\documentclass[prl,twocolumn,amsfonts,amssymb,amsmath,floatfix,tightenlines,superscriptaddress]{revtex4-1}

\usepackage{amsfonts,amssymb}
\usepackage{amsmath,times}
\usepackage{graphicx}
\usepackage{hyperref}    
\usepackage{color}
\usepackage{cleveref}
\usepackage[lofdepth,lotdepth,caption=false]{subfig}
\usepackage[normalem]{ulem}

\begin{document}

\title{Electrodynamic Signatures of Bogoliubov Fermi Surfaces in 
Moir\'e
Graphene}

\author{Ke Wang }
\email{wangke0720@gmail.com}
\affiliation{Department of Physics and James Franck Institute, University of Chicago, Chicago, Illinois 60637, USA}
\affiliation{Kadanoff Center for Theoretical Physics, University of Chicago, Chicago, Illinois 60637, USA}

\author{ Qijin Chen}
\affiliation{Hefei National Research Center for Physical Sciences at the Microscale and School of Physical Sciences, University of Science and Technology of China,  Hefei, Anhui 230026, China}
\affiliation{Shanghai Research Center for Quantum Science and CAS Center for Excellence in Quantum Information and Quantum Physics, University of  Science and Technology of China, Shanghai 201315, China}
\affiliation{Hefei National Laboratory,  Hefei 230088, China}

\author{Rufus Boyack}
\affiliation{Department of Physics and Astronomy, Dartmouth College, Hanover, New Hampshire 03755, USA}
\author{K. Levin}
\affiliation{Department of Physics and James Franck Institute, University of Chicago, Chicago, Illinois 60637, USA}

\begin{abstract}
Understanding the superfluid stiffness \(D_s\) is a central
problem in flat-band superconductivity.  It is often interpreted
together with spectroscopic probes such as tunneling, since both
ultimately reflect the same superconducting quasiparticles.
Their relationship is nevertheless complicated in flat bands,
where quantum-geometric contributions to \(D_s\) must be
included.  The situation in moir\'e graphene is particularly
complex: tunneling experiments in some cases reveal an evolution between
V- and U-shaped spectra together with a quite universal finite zero-bias conductance
(ZBC). These two phenomena along with others have
been argued to suggest the presence of a Bogoliubov Fermi
surface (BFS), often  generated by finite-momentum
pair-density-wave (PDW) superconductivity.  In this paper we calculate the superfluid
stiffness in the presence of such a PDW.
Virtual interband processes provide the familiar positive geometric
contribution that gives phase rigidity to the flat-band condensate,
allowing superconductivity to survive even in the presence of a BFS.
At the same time, the multiband PDW pair structure enables the gapless
BFS quasiparticles to respond to a phase gradient despite the negligible
ordinary flat-band velocity. Their resulting counterflow reduces the
stiffness even at zero temperature.
As an
experimentally accessible consequence, we predict a correlated
evolution of the residual ZBC and the low-temperature stiffness:
enhanced zero-bias spectral weight should accompany reduced
phase rigidity.  Observation of this correlation would support the presence of a BFS and indicate that the residual zero-energy states are intrinsic.
\end{abstract}

\date{\today}

\maketitle

\textbf{Corresponding author:} Ke Wang; \textbf{Mail address:} wangke0720@gmail.com

\section{Introduction}

The exciting discoveries surrounding superconductivity in
crystalline 
~\cite{Zhou2021,Holleis2025,Han2025}
and
moir\'e graphene
~\cite{Cao2018a,Cao2018,Yankowitz2019,Liu2020,Cao2021,caoNematic}
have called attention to the
role of flat bands which suppress the superfluid density. 
At the same time there has been a growing interest in the
role of a  
Bogoliubov Fermi surface
\cite{Agterberg2017}
(BFS) in these materials, where an extended manifold of gapless quasiparticles remains available
to similarly suppress the supercurrent response.
This introduces a fundamental question: 
can quantum geometry
~\cite{PhysRevLett.123.237002,Peotta2015}
provide sufficient stiffness to sustain phase-coherent superconductivity in a flat band system when a
Bogoliubov Fermi surface is present?

In the past a BFS has been introduced to address two experimental
features: the fact that
there is an unusual evolution in the shape of the tunneling spectrum, transitioning from
V shaped
to U shaped with stronger pairing
~\cite{Christos2023,wang2025kekule}
; quite generally this tunneling also shows~\cite{wang2025kekule} a
residual zero-bias-conductance
~\cite{Kim2022,Park2026}
(ZBC).
Different microscopic scenarios have been proposed, although here we
consider one specifically based on Kekul\'e induced pair density wave (PDW) superconductivity.

Although flat-band kinematics and a Bogoliubov Fermi surface both
undermine phase rigidity, their interplay is governed by two distinct
roles of quantum geometry.  Virtual interband processes generate a
positive contribution to the stiffness.  
For the commensurate 
~\cite{wang2025kekule}
Kekul\'e PDW, this positive contribution
stabilizes the condensate at its finite pairing wavevector and
protects it against distortions even when the ordinary band velocity
is negligible.
At the same time, the multiband structure of the PDW
pair wavefunction allows a phase gradient to shift the energies of the
gapless BFS quasiparticles.  These quasiparticles consequently generate
a paramagnetic counterflow and a negative contribution to \(D_s\) that
remains finite even at zero temperature.  Quantum geometry thus both
stabilizes the finite-momentum condensate and makes the Bogoliubov
Fermi surface electrodynamically active.


Recent stiffness measurements~\cite{Banerjee2025,Tanaka2025} in bilayer and trilayer moir\'e
graphene have not established a common low-temperature behavior, 
while tunneling experiments
~\cite{Park2026,Oh2021,Kim2022}
 have similarly
not reported a consensus characterization of the superconducting gap.
Thus, these rather direct
probes have yet to converge on a consistent picture of the
superconducting quasiparticle spectrum.
Here we argue that 
to make progress it is important to study this stiffness in conjunction with the tunneling.
Figure 1 makes this point in a schematic fashion.

\begin{figure*}
    \centering
\includegraphics[width=\linewidth]{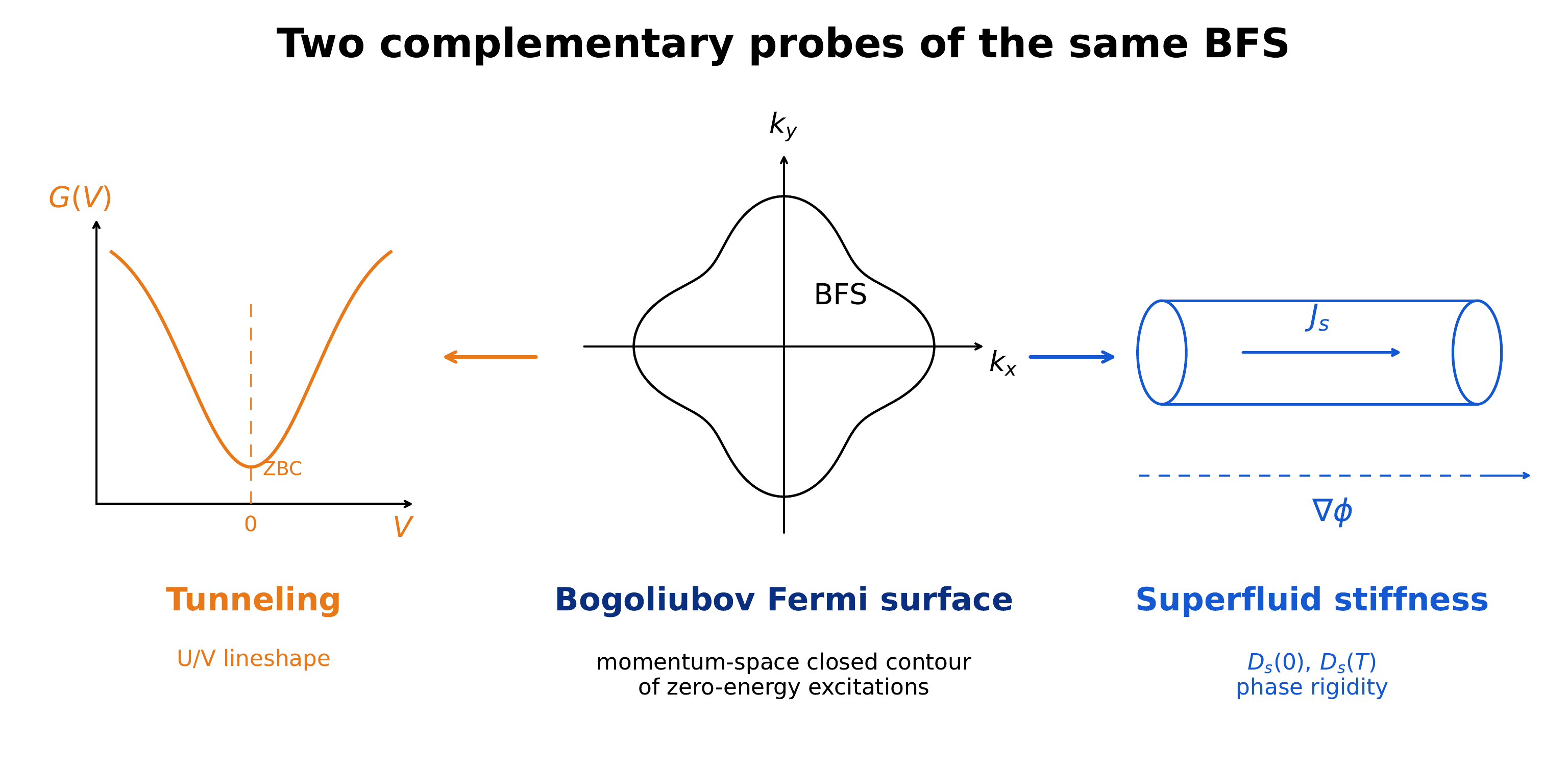}
\caption{
Two complementary probes of the same Bogoliubov Fermi surface.
Tunneling measures the residual zero-energy density of states through
the zero-bias conductance and the evolution between \(V\)- and
\(U\)-shaped spectra, while the superfluid stiffness probes the same
Bogoliubov quasiparticles weighted by their current matrix elements.
A Bogoliubov Fermi surface therefore links finite zero-bias conductance
to a reduction of the phase rigidity.
}
    \label{fig1}
\end{figure*}

The intravalley Kekul\'e pair-density-wave state considered here is
motivated by the distinct Kekul\'e modulation observed by STM~\cite{Kim2023,Nuckolls2023} in the
superconducting regime.
Particularly suggestive~\cite{Kim2023} is that the superconducting Kekulé peak intensity
is maximal
where $T_c$
 and the pairing scale are largest, rather than near an insulating phase. This correlation disfavors a 
simple inheritance from nearby normal-state order and instead points to the Kekulé modulation as an intrinsic
feature of the superconducting condensate.


The dual role of quantum geometry has experimentally accessible
consequences as we show in Figure 2. In the Bogoliubov-Fermi-surface regime 
(associated with weaker attraction) the
zero-temperature stiffness is suppressed below the
conventional quantum-geometric result and depends strongly and
nonlinearly on the pairing amplitude \(\Delta_0\). As the pairing
strength increases, the Bogoliubov Fermi surface contracts and
eventually disappears. Simultaneously, the tunneling spectrum evolves from V shaped
to U shaped~\cite{Kim2022}, and the conventional nearly linear stiffness scaling of a fully
gapped flat-band superconductor
~\cite{Peotta2015}
is reinstated. 

Also important is the fact that
tunneling measures~\cite{Kim2022,Park2026} a residual zero-energy density of states, manifested as a finite zero-bias conductance (ZBC). 
Because
stiffness probes the same states but weighted by their geometric
current matrix elements, at low temperatures the zero-bias conductance and
the superfluid stiffness should show an (anti)correlation.  A confirmation of this
evolution would support the present narrative, indicating that the zero-bias
spectral weight is an intrinsic, bulk signature of electromagnetically
active Bogoliubov quasiparticles.

Although our explicit calculations focus on magic-angle twisted
bilayer graphene, the underlying connection is more general.
Narrow-band superconductors with finite-momentum, multiband, or
valley-structured pairing can host Bogoliubov quasiparticles whose
current response is governed by pairing-specific quantum-geometric
matrix elements.  Twisted double-bilayer graphene \cite{Su2023},
alternating-twist graphene multilayers \cite{Hao2021}, rhombohedral
graphene \cite{Zhou2021}, and trilayer graphene/hBN moir\'e
superlattices \cite{Chen2019} therefore provide natural platforms in
which a similar correspondence between tunneling spectroscopy and
superfluid stiffness may arise.

\section{Bogoliubov-Fermi-Surface Contribution to Quantum Geometry and Stiffness}

The superfluid stiffness characterizes the response of a superconducting state to an electromagnetic vector potential,
\begin{equation}
\mathbf{j}=-\mathcal{D}\mathbf{A},
\end{equation}
where $\mathcal{D}$ is the response matrix $D_{ij}$
with Cartesian co-ordinates $i,j$. In an ordinary dispersive superconductor, gapless quasiparticles
couple to this phase gradient through their band velocity and reduce
the stiffness. The corresponding situation is less obvious in a
flat-band system and in the presence of a BFS, where the ordinary velocity is strongly suppressed.
The existence of a BFS alone does not establish that
its zero-energy quasiparticles remain electromagnetically active.

The superfluid stiffness also admits a useful thermodynamic
interpretation as the curvature of the grand-canonical potential with
respect to the pairing momentum,
$
D_{ij}(T)
\propto
\left.{\partial^2\Omega(\mathbf{Q},T)}
/{\partial Q_i\,\partial Q_j}
\right|_{\mathbf{Q}=\mathbf{Q}_0}.
$
We define the corresponding effective scalar stiffness as
\begin{equation}
D_s(T)\equiv \sqrt{\det[\mathcal{D}(T)]},
\end{equation}
and consider the commensurate Kekul\'e PDW which was found to be
\(\mathbf{Q}_0=\mathbf{M}\). (For a detailed discussion of this state, see the Methods Section, ``PDW Superconducting Order".)
The stiffness tensor characterizes
the rigidity of the superconducting state against deformations of its
ordering wavevector. A displacement of \(\mathbf{Q}\) along one
direction breaks the mirror relation between the two layers, whereas
a displacement along the orthogonal direction shifts the PDW away
from its moir\'e-commensurate wavevector. In the present coordinate
system, symmetry enforces \(D_{xy}=0\). Positive values of both
diagonal components, \(D_{xx}\) and \(D_{yy}\), therefore indicate
that the PDW state is locally stable against mirror-symmetry-breaking
and commensurability-breaking distortions.

The central result of this paper is that the PDW pairing form factor
provides an electrodynamic coupling. The order parameter is not simply
a scalar gap confined to the flat bands. Because the pairing form
factor $\Lambda$ is highly nonlocal in BM-band space, it contains
pairing amplitudes that connect the flat and remote bands. See Ref.~\cite{wang2025kekule} and Method Section ``PDW Superconducting Order", for a detailed discussion. These
flat--remote components combine with the Berry connection of the
normal-state wavefunctions and transfer remote-band quantum geometry
to the low-energy Bogoliubov quasiparticles. The BFS 
becomes electrodynamically active even when the ordinary
flat-band velocity is small.

\begin{figure}
    \centering
    \includegraphics[width=1\linewidth]{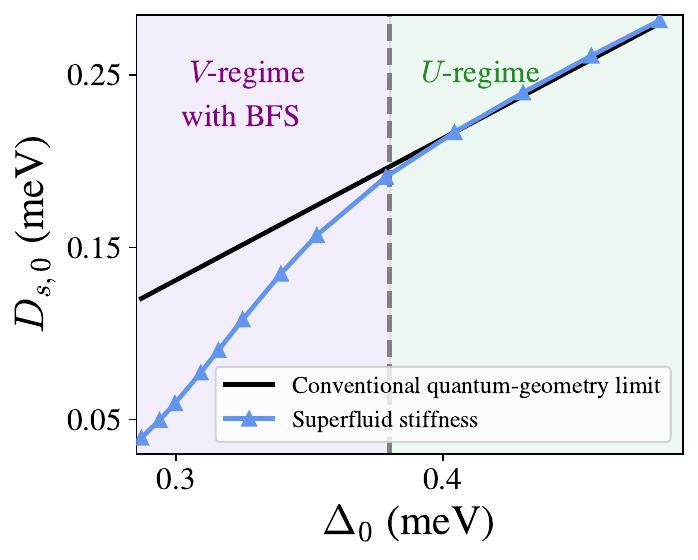}
    \caption{Zero-temperature superfluid stiffness,
\(D_{s,0}\equiv D_s(T=0)\), as a function of the
zero-temperature superconducting order-parameter amplitude
\(\Delta_0\) (blue curve). The dashed gray line indicates the crossover boundary between the $U$ and $V$ regimes. The strong pairing $U$-regime is more appropriate for tri-layer systems.
}
\label{fig2}
\end{figure}


As shown in Figure 3, we distinguish between the nearly flat and remote bands of the
normal-state BM Hamiltonian and the active and inactive bands of the
superconducting BdG Hamiltonian. Because the PDW pairing form factor
mixes BM bands, an active BdG quasiparticle generally contains both
flat- and remote-band components.
As shown previously~\cite{wang2025kekule}, dominant inter-flat-band pairing divides the
low-energy BdG Hamiltonian into two approximately decoupled sectors.
The mismatch between the paired dispersions allows a Bogoliubov branch
to cross zero energy along an extended contour, producing a robust
BFS~\cite{wang2025kekule}. Here we focus on how these gapless
quasiparticles couple to a phase gradient.

Quantum geometry
enters this response in two distinct ways. The first
is the familiar contribution 
present in conventional flat-band
theories~\cite{PhysRevLett.123.237002,Peotta2015}. Virtual processes
between active and inactive BdG sectors provide a positive geometric
background to the stiffness, allowing the condensate to retain finite
phase rigidity even when the ordinary kinetic contribution is small.

The PDW pairing form factor generates an additional,
pairing-dependent geometric channel entirely within the active BdG
sector. Through flat--remote-band pairing and the Berry connection,
this channel makes the gapless Bogoliubov-Fermi-surface quasiparticles
electrodynamically active. Their response is paramagnetic and
therefore reduces the stiffness, producing a negative contribution
that survives even at zero temperature. Thus, quantum geometry not
only sustains the flat-band condensate, but also makes the Bogoliubov
Fermi surface responsive electrodynamically.



\begin{figure*}
    \centering
    \includegraphics[width=.8\linewidth]{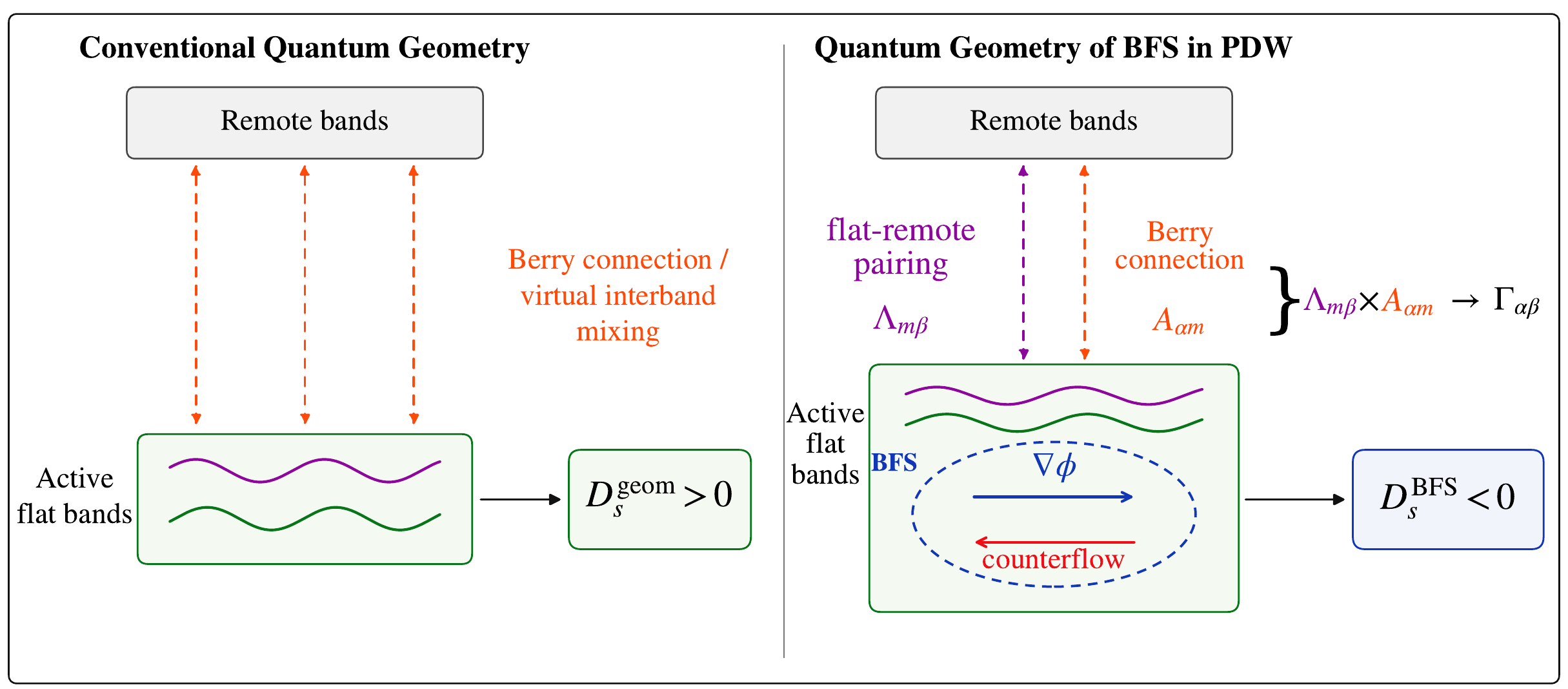}
    \caption{ \textbf{Conventional quantum geometry versus Bogoliubov
        Fermi-surface geometry in a pair-density-wave state.}
        \textbf{Left:} Virtual interband mixing with remote bands produces
        a positive geometric contribution to the superfluid stiffness,
        $D_{ii}^{\mathrm{geom}}>0$ with $i=x,y$.
        \textbf{Right:} Flat--remote pairing, $\Lambda_{m\beta}$, combines
        with the Berry connection, $A_{\alpha m}$, to form the current
        vertex $\Gamma_{\alpha\beta}$. The resulting Bogoliubov Fermi
        surface (BFS) supports phase-gradient-induced counterflow and
        contributes negatively to the stiffness,
        $D_{ii}^{\mathrm{BFS}}=(\mathcal{D}_{\mathrm{BFS}})_{ii}<0$ with $i=x,y$.
}
\label{fig3}
\end{figure*}

The geometric origin of this coupling can be made explicit by
introducing the matrix
\begin{equation}
\Gamma_{\alpha\beta}(\mathbf{k})
=
i\sum_{m\in\mathrm{remote}}
\left[
\mathcal{A}_{\alpha m}(\mathbf{k})\Lambda^{*}_{m\beta}(\mathbf{k})
-
\mathcal{A}_{m\alpha}(\mathbf{k})\Lambda_{m\beta}(\mathbf{k})
\right],
\label{eq:2}
\end{equation}
where
\begin{equation}
\mathcal{A}_{mn}(\mathbf{k})
=
i\langle u_m(\mathbf{k})|\nabla_{\mathbf{k}}u_n(\mathbf{k})\rangle
\end{equation}
is the Berry connection of the BM bands. The quantity
$\Gamma_{\alpha\beta}$ combines the Cooper-pair wavefunction with the
normal-state Berry geometry. Note that $\Gamma_{\alpha\beta}(\mathbf{k})$ is invariant under gauge transformations of the remote-band states,
$\vert m\rangle\to e^{i\alpha_m(\mathbf{k})}\vert m\rangle$.
In this sense, $\Lambda$ provides the
coupling to the remote bands, while $A$ supplies the geometric current
matrix element. Neither ingredient alone produces the full effect.

\begin{figure*}
    \centering
    \includegraphics[width=.8\linewidth]{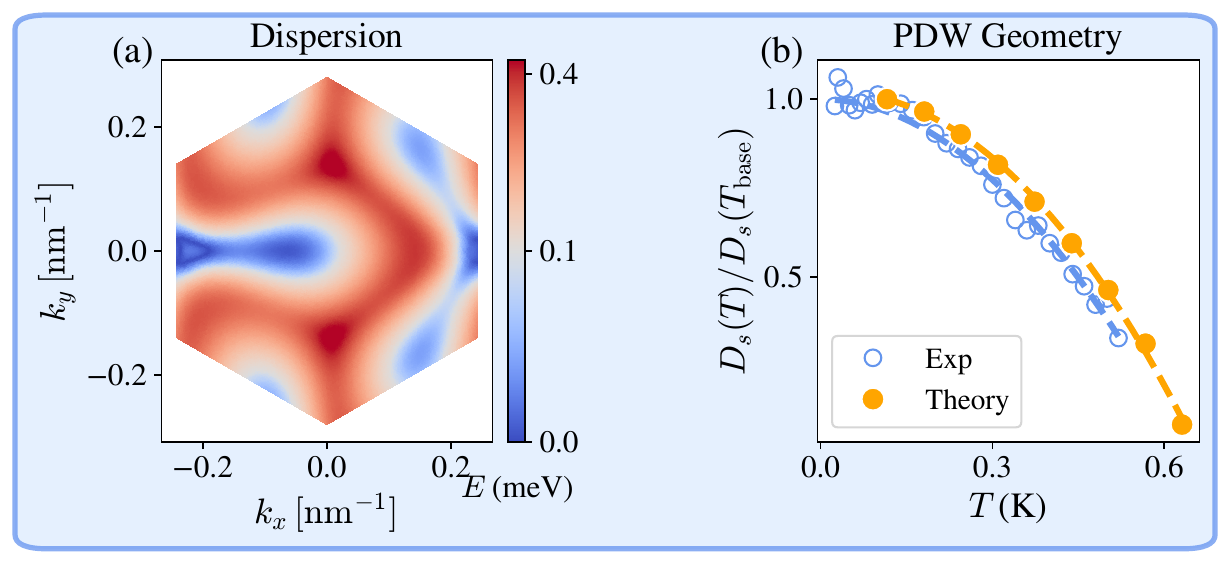}
    \caption{
(a) Lowest nonnegative BdG eigenvalue in the active band at the order parameter $\Delta=0.31\,\mathrm{meV}$, showing a
small gapless Bogoliubov Fermi surface (in dark blue). (b) Comparison between theory and experiment. The experimental data are adapted from Fig.~3(a) of Ref.~\cite{Tanaka2025}, which shows the normalized superfluid stiffness in the hole-doped regime. In the theoretical curve, \(D_{s}(T)\) is computed at $\Delta=0.31\,\mathrm{meV}$, with reference temperature $T_{\mathrm{base}}\simeq 0.1\,\mathrm{K}$. The dashed lines denote power-law fits: the theoretical result gives an exponent $n\simeq 2.1$, while the experimental data are consistent with $n\simeq 2$. 
}
    \label{fig4}
\end{figure*}

The resulting BFS contribution to the stiffness can be written as
\begin{equation}
\mathcal{D}_{\mathrm{BFS}}(T)
=
4\int dE\,n_F'(E)\mathcal{W}(E),
\label{DBFS}
\end{equation}
where $n_F'(E)$ is the derivative of the Fermi function and
$\mathcal{W}(E)$ is a geometrically weighted density of states,
\begin{equation}
\mathcal{W}(E)
=
\sum_{\alpha,s,\mathbf{k}}
\delta\!\left(E-E_{\alpha,s}(\mathbf{k})\right)
\left[
\frac{
\Lambda_{\alpha\beta}(\mathbf{k})\Delta^2
}{
2\epsilon_\alpha(\mathbf{k})
}
\right]^2
\Gamma_{\alpha\beta}(\mathbf{k})
\Gamma_{\beta\alpha}(-\mathbf{k}).
\label{WeightedDOS}
\end{equation}
Here $\alpha=1,2$, $s=\pm$, and $\beta\neq\alpha$ 
denotes the other
flat band. The derivation is provided in the Methods section, “Geometric BFS”. The quantity $\mathcal{W}$ is invariant under gauge transformations of the flat-band states due to the presence of $\Lambda$ in Eq.~\ref{WeightedDOS}.

Equation~\ref{DBFS} expresses the main theoretical result. The
low-energy density of states is not sufficient by itself to determine
the electromagnetic response. Each Bogoliubov state is weighted by a
current matrix element that depends on both the multiband Cooper-pair
wavefunction and the Berry connection. Because these matrix elements
remain finite as the flat-band dispersion is reduced, the BFS
contribution is not parametrically suppressed in the narrow-band
limit. Although Ref.~\onlinecite{Putzer2025} captures quantum-geometric
effects within an effective two-flat-band model, the specific
Bogoliubov-Fermi-surface contribution in
Eqs.~\eqref{DBFS} and \eqref{WeightedDOS} is absent from that
treatment. Obtaining this contribution explicitly requires retaining
the remote-band structure of the microscopic BM model, since it
arises from flat--remote-band pairing components combined with the
corresponding Berry-connection matrix elements.



\begin{figure*}
    \centering
    \includegraphics[width=.8\linewidth]{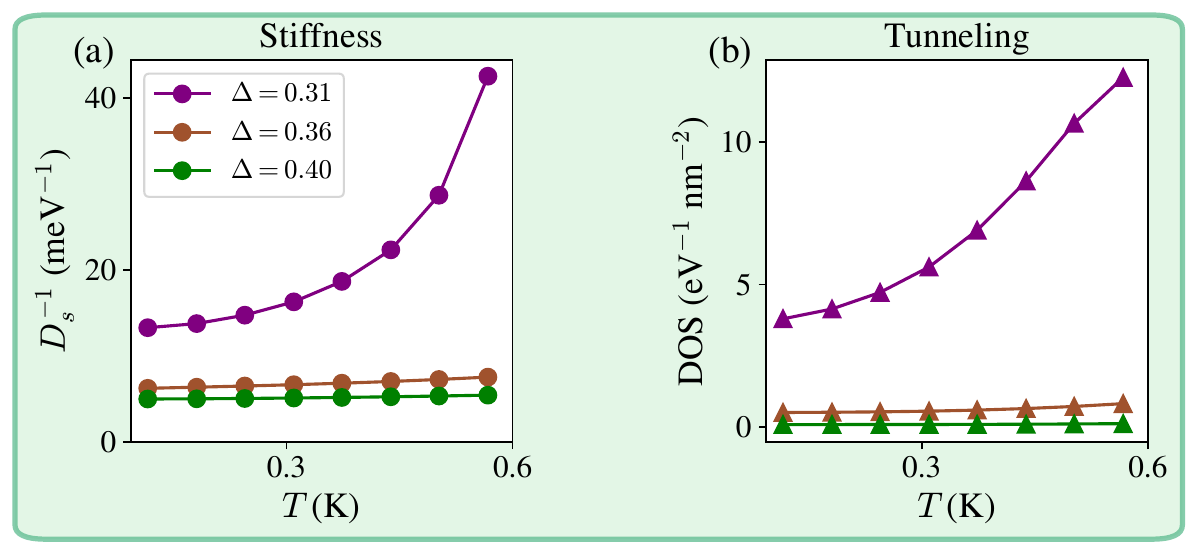}
    \caption{
(a) Temperature dependence of the inverse bare superfluid stiffness, $D_s^{-1}(T)$, for representative gap values $|\Delta|=0.40$, $0.36$, and $0.31\,\mathrm{meV}$. (b) Corresponding zero-energy density of states, $\mathrm{DOS}(E=0)$, which is proportional to the zero-bias tunneling conductance, for the same parameters. Over the superconducting temperature range shown, $D_s^{-1}(T)$ closely tracks the evolution of $\mathrm{DOS}(E=0)$.}
  \label{fig5}
\end{figure*}

\section{Finite-Temperature Stiffness and the Tunneling Connection}

Figure~4(a) reveals the Bogoliubov Fermi surface explicitly by
mapping the lowest nonnegative BdG eigenvalue,
\(E_{\min}(\mathbf{k})\), throughout the moir\'e Brillouin zone.
In a fully gapped superconductor, \(E_{\min}(\mathbf{k})\) remains
positive at every momentum. For the representative value
\(\Delta=0.31\,{\rm meV}\), it instead vanishes in the dark-blue
region, identifying the Bogoliubov Fermi surface.

Having established explicitly the presence of a BFS in Fig.~4(a), we now turn to the temperature dependence of the stiffness in Fig.~4(b). We find that in
the $V$-shaped regime, the
low-temperature stiffness is well described by
\begin{equation}
D_s(T)
\approx
D_s(0)-aT^n,
\end{equation}
with
\begin{equation}
n=2\pm0.2.
\end{equation}
A representative result yields $n\simeq2.1$, as shown in Fig.~4b. 
Deviations from a perfectly quadratic
law arise because $\Delta(T)$ changes the size of the BFS and hence
the available low-energy phase space. Once this variation becomes
appreciable, the strict asymptotic expansion in
Eq.~\ref{LowTStiffness} is no longer sufficient.

To understand the origin of this power law, note that once
the BFS quasiparticles are able to participate in electrodynamics
of the condensate it follows that their thermal occupation directly controls the
low-temperature stiffness. The temperature dependence of $D_s(T)$
therefore probes not only the existence of low-energy quasiparticles
but also their coupling to the condensate.

We might expect that in the asymptotically low-temperature regime, where $T$ is much
smaller than the other relevant energy scales, the stiffness is a consequence
of a Sommerfeld-like expansion, taking
the form
\begin{equation}
D_s(T)
\simeq
D_s(0)-aT^2.
\label{LowTStiffness}
\end{equation}
The coefficient $a$ is not determined solely by the ordinary density
of states. It contains the geometric weighting of the BFS, for
example through derivatives of $\mathcal{W}(E)$ evaluated near
$E=0$.


The near-\(T^2\) behavior is consistent with experiment~\cite{Tanaka2025}, but the
exponent alone is not a unique signature of a BFS. A more decisive
test can be found by comparing tunneling and stiffness under the same
tuning conditions.
Tunneling measures
the ordinary quasiparticle density of states,
\begin{equation}
N(E)
=
\sum_{\alpha,s,\mathbf{k}}
\delta\!\left(E-E_{\alpha,s}(\mathbf{k})\right),
\end{equation}
so that the residual zero-bias conductance is proportional to $N(0)$.
By contrast, the stiffness probes the same states with the additional
geometric weight contained in $\mathcal{W}(E)$.

At zero temperature,
\begin{equation}
\mathcal{D}_{\mathrm{BFS}}(0)
=
-4\mathcal{W}(0).
\end{equation}
In this way, the residual zero-bias states detected by tunneling are connected to the quasiparticle contribution to the stiffness. This is the direct electrodynamic consequence of the flat--remote structure of the PDW pairing form factor.

Importantly, this analysis predicts a correlated evolution of tunneling and
stiffness. The larger the BFS, the greater the residual zero-bias conductance. The corresponding enhancement of the negative quasiparticle contribution reduces the low-temperature stiffness. Conversely,
as the BFS contracts and disappears across the $V$-to-$U$ crossover,
the residual zero-bias spectral weight will decrease and the
stiffness will recover.

Because tunneling measures $N(0)$ whereas the stiffness depends on the
geometrically weighted quantity $\mathcal{W}(0)$, the two observables
are not expected to obey a universal proportionality. Their
correlated evolution
is the
central prediction we present in Figure 5.
Over the superconducting temperature range shown, 
we can see that $D_s^{-1}(T)$ closely tracks the evolution of the ZBC or $\mathrm{DOS}(E=0)$.

This comparison importantly tests whether the residual zero-bias spectral
weight is intrinsic. Phenomenological Dynes broadening, instrumental
resolution, or localized interface states can produce finite
zero-bias conductance without any necessary change in the bulk
superfluid stiffness. By contrast, a correlated suppression of $D_s$
would show that the residual states are bulk Bogoliubov quasiparticles
that participate in the electrodynamic response of the condensate.

\begin{figure}
    \centering
    \includegraphics[width=0.9\linewidth]{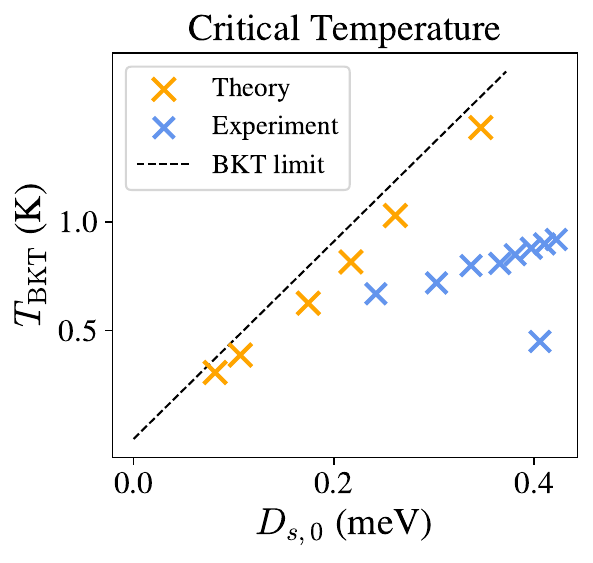}
    \caption{ Plot of the BKT transition temperature $T_{\text{BKT}}$ versus the zero-temperature superfluid stiffness, motivated by Fig.~2h in Ref.~\cite{Tanaka2025}. The experimental data are adapted from the hole-doped data in Fig.~2h of Ref.~\cite{Tanaka2025}, using the authors' definition of $T_c^{(0.5)}$ for points with $D_{s,0}\leq 0.42$ meV.
}
    \label{fig6}
\end{figure}


Finally, in Figure 6 we calibrate some of these observations by comparing
the calculated BKT temperature directly with experiment.
For the superconducting state considered here, we find an approximately linear relation between $T_{\rm BKT}$ and $\pi D_{s,0}$ over a broad range of vortex energies, $E_v$. The corresponding slope, $\eta=T_{\rm BKT}/\pi D_{s,0}$, is nevertheless nonuniversal and depends on microscopic details such as the flat-band structure and the vortex-core energy. Figure 6 shows a representative result for the canonical choice $E_v=\pi\rho_s$, together with the experimental trend 
relating $T_{\rm BKT}$ and $D_s$. These data are adapted\footnote{Here we use the data for carrier densities below $1.5\times10^{11}\,\mathrm{cm}^{-2}$ in Fig.~2h of Ref.~\cite{Tanaka2025}.} from Fig.~2h of Ref.~\cite{Tanaka2025}.








\section{Conclusions}

Understanding the superfluid stiffness is a central problem in
flat-band superconductivity, where the ordinary velocity contribution
to the supercurrent is strongly suppressed. The problem becomes still
more severe when the superconducting state contains a Bogoliubov
Fermi surface: an extended manifold of zero-energy quasiparticles
remains available to oppose the condensate flow. This raises a basic
question: can a flat-band superconductor retain phase rigidity without
eliminating its Bogoliubov Fermi surface?

This paper shows that it can, but through a more intricate role of quantum
geometry than in conventional flat-band theories. 
Virtual interband processes provide the familiar positive geometric
contribution that gives the condensate finite stiffness. At the same
time, the multiband structure of the PDW pair wavefunction allows a
phase gradient to shift the energies of the gapless Bogoliubov
quasiparticles even when the ordinary flat-band velocity is
negligible. These quasiparticles therefore participate in the
supercurrent response without becoming ordinary particles of definite
charge. Their current opposes the condensate flow and produces a
negative contribution to \(D_s\) that remains finite even at zero
temperature. Quantum geometry therefore both sustains the condensate
and determines how the Bogoliubov Fermi surface weakens its phase
rigidity.

This dual role leads to a direct experimental prediction. Tunneling
measures the ordinary zero-energy quasiparticle density of states,
whereas the superfluid stiffness probes the same states weighted by
their quantum-geometric current matrix elements. Tuning that enhances
the residual zero-bias conductance should therefore reduce the
low-temperature stiffness. Conversely, disappearance of the residual
zero-bias weight across the V-to-U crossover should be accompanied by
recovery of the stiffness. This linked evolution would distinguish
intrinsic bulk Bogoliubov quasiparticles from phenomenological
tunneling broadening or localized interface states.

\section{Methods}

\subsection{ PDW Superconducting Order} 

We briefly summarize the PDW/Kekulé superconducting state established in Ref.~\cite{wang2025kekule}. We begin with the Bistritzer--MacDonald (BM) continuum model supplemented by a short-range attractive interaction that preserves the symmetries of the BM Hamiltonian and couples the $A$ and $B$ sublattices as for a nearest-neighbor intersite attraction. The resulting superconducting state is characterized by PDW condensates on the $AB$ valence bonds, $\Delta_L(2\mathbf{Q})$ and $\Delta_L(-2\mathbf{Q})$, where $2\mathbf{Q}$ denotes the Cooper-pair center-of-mass momentum and $L$ is the layer index. The condensates in the two valleys are related by time-reversal symmetry $\mathcal{T}$ and inversion symmetry $\mathcal{C}_2$. Owing to the large momentum separation between valleys, we treat them as effectively decoupled, with the opposite-valley physics obtained by applying $\mathcal{T}$. Thus valley symmetry $U(1)_v$ is preserved.

Minimization of the grand-canonical thermodynamic potential $\Omega(\bf{Q})$, with the chemical potential tuned near the flat-band regime, yields an intravalley unitary spin-triplet state with ordering wave vector $\mathbf{Q}=\mathbf{M}$, in which the condensates on the two layers are equal. Here $\mathbf{M}$ denotes the $\mathbf{M}$ point of the mini Brillouin zone, located midway between the Dirac momenta of the top and bottom layers, $\mathbf{K}_L$. Consequently, this state spontaneously breaks the $\mathcal{C}_3$ rotational symmetry while preserving the mirror symmetry $\mathcal{M}$.
Since the condensate momentum $2\mathbf{M}\in \mathcal{G}$ belongs to the moiré reciprocal lattice, the PDW state is compatible with moiré translation symmetry.
Note that the induced charge-density-wave order exhibits a rapid atomic-scale Kekul\'e modulation\cite{Nuckolls2023}, together with a slowly varying moir\'e-scale envelope at momentum $4(\mathbf{M}-\mathbf{K}_L)$ superimposed on the Kekul\'e pattern\cite{wang2025kekule}.
Although the full PDW state preserves $\mathcal{T}$ and  $\mathcal{C}_2$, the pairing process occurs within a single valley, where neither $\mathcal{T}$ nor $\mathcal{C}_2$ is individually present. This feature underlies much of the rich phenomenology of the superconducting state. 

To understand the consequences of this symmetry structure, we focus on the particle-particle form factor $\Lambda$, which characterizes the Cooper-pair wavefunction in band space and over the mini Brillouin zone. Schematically,
$
\Lambda_{mn}(\mathbf{q}) \sim \sum_{\mathbf{G}} u_{\mathbf{G};m}(\mathbf{Q}+\mathbf{q})\,u_{-\mathbf{G};n}(\mathbf{Q}-\mathbf{q})
$ \footnote{In the full expression, one must also include the symmetry factors associated with the short-range interaction, as well as microscopic indices such as layer and sublattice.},
where $u_{\mathbf{G};m}(\mathbf{q})$ is the BM-model Bloch wavefunction of band $m$ at reciprocal-lattice component $\mathbf{G}$, and $\mathbf{G}$ runs over moir\'e reciprocal lattice vectors. If $\mathcal{T}$ and $\mathcal{C}_2$ were present, $\Lambda$ would be reduced to an overlap of BM wavefunctions and would therefore be proportional to the identity in band space and independent of $\mathbf{q}$. Here, however, neither symmetry is individually present, so $\Lambda$ becomes strongly momentum dependent and highly nonlocal in band space. 

Importantly, the inter-flat-band form factor $\Lambda_{12}$ is sizable and plays an important role in stabilizing a unitary spin-triplet state over a spin-singlet state, while the diagonal pairing form factor $\Lambda_{11}$ is strongly suppressed for the triplet channel by $\mathcal{C}_2\mathcal{T}$ symmetry. Here $1$ and $2$ label the two flat bands, with flat-band energies $\xi_1<\xi_2$.

\subsection{Physical States and Superfluid Stiffness}
We focus on the Kekul\'e superconducting state, which is characterized by two condensates at pairing momenta $\pm 2\mathbf Q$. Owing to the large momentum separation between the two Dirac cones, we treat the two valleys as effectively decoupled. The thermodynamic potential of the full system is then obtained by summing the contributions from the two valleys.

Given the BM band structure and the attractive interaction, we compute the grand-canonical thermodynamic potential for a general mean-field ansatz,
\begin{equation}
\Omega(\mu,\mathbf Q,\Delta),
\end{equation}
at zero temperature. 
For a fixed chemical potential $\mu$, we determine the thermodynamically stable state in two steps. First, for given $(\mu,\mathbf Q)$ we minimize $\Omega$ with respect to $\Delta$, which yields the self-consistent gap $\Delta_0(\mu,\mathbf Q)$ satisfying
\begin{equation}
\left.\frac{\partial \Omega}{\partial \Delta}\right|_{\mu,\mathbf Q}=0,
\qquad
\left.\frac{\partial^2 \Omega}{\partial \Delta^2}\right|_{\mu,\mathbf Q}>0.
\end{equation}
This defines the on-shell potential
\begin{equation}
\Omega_0(\mu,\mathbf Q)\equiv \Omega\!\big(\mu,\mathbf Q,\Delta_0(\mu,\mathbf Q)\big).
\end{equation}
Second, we minimize $\Omega_0$ with respect to $\mathbf Q$ over the mini-Brillouin zone, i.e.
\begin{equation}
\left.\frac{\partial \Omega_0}{\partial Q_i}\right|_{\mu}=0,
\qquad
 \frac{\partial^2 \Omega_0}{\partial Q_i\partial Q_j} \text{is positive definite},
\end{equation}
which determines the optimal momentum $\mathbf Q(\mu)$. Numerically, scanning a range of $\mu$ we find that $\mathbf Q(\mu)$ is pinned to the $M$ point of the mini-Brillouin zone, consistent with spontaneous $C_3$ symmetry breaking. The resulting thermodynamic potential is
\begin{equation}
\Omega_1(\mu)=\Omega_0\!\big(\mu,\mathbf Q(\mu)\big).
\end{equation}

Within this mean-field framework, the (bare) superfluid stiffness is given by the curvature of the on-shell grand potential with respect to $\mathbf Q$,
\begin{equation}
D_{ij}
=\frac{\partial^2 \Omega_0(\mu,\mathbf Q)}{\partial Q_i\,\partial Q_j}
=\frac{d^2}{dQ_i\,dQ_j}\,\Omega\!\big(\mu,\mathbf Q,\Delta_0(\mu,\mathbf Q)\big),
\label{S4}
\end{equation}
reflecting the equivalence between a phase twist and coupling to a uniform vector potential, $\partial/\partial \mathbf Q \leftrightarrow \partial/\partial \mathbf A$. We emphasize that Eq.~\eqref{S4} applies to the present Kekul\'e state because the $\pm2\mathbf Q$ condensates are effectively independent (a tensor-product structure). For more general PDW states in which the $\pm2\mathbf Q$ sectors are coupled, the stiffness cannot be obtained from a simple curvature of $\Omega_0$; instead one must compute the full current--current response including the appropriate vertex corrections (see, e.g., Ref.~\cite{wang2025}). Thus, for the present Kekul\'e state, the superfluid stiffness is directly the curvature of the thermodynamic potential with respect to the pairing momentum, providing a measure of the local rigidity of the PDW state.

\subsection{Ensemble Dependence and Stability}
An analogous construction applies in the canonical ensemble, where one works with the free energy $F(N,\mathbf Q,\Delta)$. Following the same procedure, one determines the self-consistent gap $\Delta_0(N,\mathbf Q)$ and then minimizes with respect to $\mathbf Q$ to obtain $\mathbf Q_0(N)$. The corresponding on-shell free energy is
\begin{equation}
F_0(N,\mathbf Q)\equiv F\!\big(N,\mathbf Q,\Delta_0(N,\mathbf Q)\big).
\end{equation}

The stationarity conditions obtained from $\Omega$ and $F$ are equivalent: the same $\Delta_0$ solves the gap equation in both ensembles, and the extremum in $\mathbf Q$ occurs at the same $\mathbf Q_0$ when evaluated at fixed $\mu$ (grand canonical) or fixed $N$ (canonical). More explicitly,
\begin{equation}
\left.\frac{\partial F}{\partial \Delta}\right|_{N,\mathbf Q}
=
\left.\frac{\partial \Omega}{\partial \Delta}\right|_{\mu,\mathbf Q},
\qquad
\left.\frac{\partial F_0}{\partial Q_i}\right|_{N}
=
\left.\frac{\partial \Omega_0}{\partial Q_i}\right|_{\mu}.
\end{equation}
Thermodynamic stability requires the relevant curvatures in both ensembles to be positive. However, the \emph{values} of these curvatures are not generally the same; in fact, they typically differ.

To illustrate this point, consider the curvature with respect to $\Delta$ in the simplest setting where $\mathbf Q$ is fixed (and suppressed in the notation), i.e., $F(N,\Delta)$ and $\Omega(\mu,\Delta)$. Using $F=\Omega+\mu N$ and taking derivatives at fixed $N$, one finds
\begin{equation}
\left.\frac{\partial^2 F(N,\Delta)}{\partial \Delta^2}\right|_{N}
=
\left.\frac{\partial^2 \Omega(\mu,\Delta)}{\partial \Delta^2}\right|_{\mu}
+
\left.\frac{\partial^2 \Omega(\mu,\Delta)}{\partial \Delta\,\partial \mu}\right|
\left.\frac{\partial \mu}{\partial \Delta}\right|_{N},
\label{S9}
\end{equation}
where $\mu=\mu(N,\Delta)$ is determined implicitly by the number equation
$N=-\left.\partial\Omega/\partial\mu\right|_{\Delta}$.
In obtaining Eq.~\eqref{S9} we used the identity
\begin{equation}
\left.\frac{\partial (\Omega+\mu N)}{\partial \Delta}\right|_{N}
=
\left.\frac{\partial \Omega(\mu,\Delta)}{\partial \Delta}\right|_{\mu}.
\end{equation}
Differentiating the number equation at fixed $N$ yields
\begin{equation}
\frac{\partial}{\partial\Delta}\left( N+\frac{\partial \Omega}{\partial \mu} \right)=0
\;\Longleftrightarrow\;
\Omega_{\mu\mu}\left.\frac{\partial \mu}{\partial \Delta}\right|_{N}
+\Omega_{\mu\Delta}=0,
\label{S11}
\end{equation}
where subscripts denote partial derivatives.
Combining Eqs.~\eqref{S9} and \eqref{S11} gives the standard relation
\begin{equation}
F_{\Delta\Delta}
=
\Omega_{\Delta\Delta}
-\frac{\Omega_{\mu\Delta}^2}{\Omega_{\mu\mu}}.
\end{equation}
Using $\Omega_{\mu\mu}=-(\partial N/\partial\mu)_{\Delta}$, we note that stability requires a positive compressibility $(\partial N/\partial\mu)_{\Delta}>0$, hence $\Omega_{\mu\mu}<0$, and therefore
\begin{equation}
F_{\Delta\Delta}>\Omega_{\Delta\Delta}.
\end{equation}
Thus, even though the stationary points coincide, the curvatures in the two ensembles are generically different.

The same logic applies to the curvature with respect to $\mathbf Q$. One may repeat the derivation above with $F(N,\Delta)\to F_0(N,\mathbf Q)$ and $\Omega(\mu,\Delta)\to \Omega_0(\mu,\mathbf Q)$, leading to an analogous relation between the $\mathbf Q$-curvatures in the canonical and grand-canonical ensembles. Thus, while the stationary superconducting state is the same in the canonical and grand-canonical
ensembles, its curvature, and hence the quantitative measure of stability, is in general ensemble
dependent.

\subsection{Diamagnetic Terms}
For the superfluid stiffness computed within linear response, it is convenient to separate the result into diamagnetic and paramagnetic contributions. The paramagnetic part is the standard current--current correlation function. Here we focus on the diamagnetic term,
\begin{equation}
[\mathrm{Dia}] \equiv
\frac{1}{2}\sum_i n_F(E_i)\,\langle i|\partial_\mu\partial_\nu\!\left(H_{\mathrm{BM}}\tau_z\right)|i\rangle,
\end{equation}
where $|i\rangle$ and $E_i$ are the eigenstates and eigenvalues of the BdG Hamiltonian $H$, and $H_{\mathrm{BM}}\tau_z$ denotes the diagonal (normal) block of $H$ in Nambu space.

One may rewrite the matrix element as
\begin{align}
& \langle i|\partial_\mu\partial_\nu
(H_{\mathrm{BM}}\tau_z)|i\rangle
={}
\partial_\mu
\langle i|\partial_\nu(H_{\mathrm{BM}}\tau_z)|i\rangle
\nonumber\\
&-\langle \partial_\mu i|
\partial_\nu(H_{\mathrm{BM}}\tau_z)|i\rangle -\langle i|
\partial_\nu(H_{\mathrm{BM}}\tau_z)
|\partial_\mu i\rangle .
\end{align}
For compactness, we define
$
V_\nu \equiv \partial_\nu(H_{\mathrm{BM}}\tau_z).$
Using completeness and standard perturbation theory for the BdG
eigenstates, we obtain
\begin{align}
\langle \partial_\mu i|V_\nu|i\rangle
&=
\sum_{j\neq i}
\frac{\langle i|\partial_\mu H|j\rangle}
     {E_i-E_j}
\langle j|V_\nu|i\rangle ,
\label{eq:deriv1}\\
\langle i|V_\nu|\partial_\mu i\rangle
&=
\sum_{j\neq i}
\langle i|V_\nu|j\rangle
\frac{\langle j|\partial_\mu H|i\rangle}
     {E_i-E_j}.
\label{eq:deriv2}
\end{align}
Here \(H\) is the full BdG Hamiltonian. Collecting terms gives
\begin{align}
\langle i|
\partial_\mu\partial_\nu
(H_{\mathrm{BM}}\tau_z)
|i\rangle
\nonumber =& 
\partial_\mu\langle i|V_\nu|i\rangle
 -\sum_{j\neq i}
\frac{
\langle i|\partial_\mu H|j\rangle
\langle j|V_\nu|i\rangle
}{
E_i-E_j
}
\nonumber\\
&\qquad
-\sum_{j\neq i}
\frac{
\langle i|V_\nu|j\rangle
\langle j|\partial_\mu H|i\rangle
}{
E_i-E_j
}.
\label{eq:second_derivative}
\end{align}
The first term can be combined with the Fermi factor by integration
by parts:
\begin{align}
n_F(E_i)\,
\partial_\mu\langle i|V_\nu|i\rangle
&=
-n_F'(E_i)
\langle i|\partial_\mu H|i\rangle
\langle i|V_\nu|i\rangle
+\text{total-derivative} .
\label{eq:fermi_integration}
\end{align}
The total-derivative term vanishes after integration over the
Brillouin zone. The diamagnetic contribution can therefore be written
as
\begin{align}
&\frac{1}{2}
\sum_i n_F(E_i)
\langle i|
\partial_\mu\partial_\nu
(H_{\mathrm{BM}}\tau_z)
|i\rangle
\nonumber\\
&\quad =
-\frac{1}{2}
\sum_{i,j}
\frac{n_F(E_i)-n_F(E_j)}
     {E_i-E_j}
\langle i|\partial_\mu H|j\rangle
\langle j|V_\nu|i\rangle .
\label{eq:dia_compact}
\end{align}
For \(i=j\), the ratio is understood in the limiting sense,
\begin{equation}
\lim_{E_j\to E_i}
\frac{n_F(E_i)-n_F(E_j)}
     {E_i-E_j}
=
n_F'(E_i).
\end{equation}

This identity is the key starting point for evaluating the
diamagnetic term in the superconducting state. If the pairing matrix
\(\Delta(\mathbf{k})\) in the microscopic orbital/layer/sublattice
basis has appreciable momentum dependence, its derivatives must be
retained in \(\partial_\mu H\). When combined with the paramagnetic
contribution, the normal-state \(GG\) response cancels, leaving only
terms that require a nonzero order parameter. In the present
Kekul\'e state, \(\Delta(\mathbf{k})\) is only weakly momentum
dependent, and in the main text we therefore focus on the anomalous
(\(FF\)) contribution.

\subsection{Geometric Active Bands}
In the main text, we mentioned that the contribution of the active bands to the
superfluid stiffness is geometric. In this section, we provide a detailed
derivation. To see why the active bands carry geometric contributions, consider
the expectation value of the current operator in the particle sector of the BdG
bands:
\begin{equation}
 J^p_{ij}(k)
 =  \langle \varphi^p_i(k) | J(\mathbf{k}) |\varphi^p_j(k) \rangle
 =\sum_{mn} [\varphi^p_i(k)]^*_n\, J_{nm}\,  [\varphi^p_j(k)]_m .
\end{equation}

We now take $i,j$ to belong to the active BdG bands. Due to the
off-diagonal structure of the pairing form factor, the wave function is spread
over the band basis; in particular, the index $n$ can run over remote bands.
Recall that $J_{\alpha m}$ is proportional to the Berry connection
$A_{\alpha m}$. As a result, the current matrix elements within the active-band
sector inherit geometric contributions from the non-Abelian Berry connections. Below we present a more systematic derivation.

\subsection{Wave Function of the Active Bands}
The first step is to derive the wave functions of the active bands. Recall that
the mean-field Hamiltonian is
\begin{align}
H_{\mathrm{MF}}
={}&
\sum_{q,n,\sigma}
\xi_n(q)\,
c^\dagger_{n,\sigma}(q)c_{n,\sigma}(q)
 \\
&-
\sum_{q,m,n,\sigma,\sigma'}
\Delta_{\sigma\sigma'}\,
\Lambda^{mn*}(q)\,
c^\dagger_{\sigma,m}(q)
c^\dagger_{\sigma',n}(-q)
+\mathrm{H.c.}\nonumber
.
\end{align}
Here $m$ labels the BM band and $\sigma$ denotes the spin. We now exploit the
large energy separation between the flat bands and the other (remote) bands in
the BM model. Since we work with the BdG Hamiltonian, we partition the
particle--hole doubled band space into low-energy (flat) and high-energy
(remote) sectors, denoted by $L$ and $H$, respectively. The BdG eigenstate then
satisfies
\begin{equation}
    H_{LL}\psi_L + H_{LH}\psi_H = E\psi_{L},\quad 
    H_{HL}\psi_L + H_{HH}\psi_H = E\psi_{H}.
\end{equation}
For the active BdG bands, we expect the dominant weight to lie in the $L$
sector, while the high-energy sector provides only a perturbative correction.
Solving the coupled equations gives
\begin{eqnarray}
\psi_H=-(H_{HH}-E)^{-1}H_{HL}\psi_L. \label{S23}
\end{eqnarray}
When $E$ is much smaller than the eigenvalues of $H_{HH}$, we approximate
\begin{eqnarray}
(H_{HH}-E)^{-1}\Big|_{m}\simeq 
\begin{pmatrix}
    \xi_m(k) & 0 \\
    0 & -\xi_m(-k)
\end{pmatrix}^{-1}. \label{S24}
\end{eqnarray}
Here the subscript $m$ indicates the BdG block associated with BM band $m$, and
$\xi_m$ is the corresponding BM band energy. Note that neither the
quasiparticle energy $E$ nor the order parameter $\Delta$ appears in Eq.~\ref{S24}. The block $H_{HL}$ reads
\begin{eqnarray}
\label{S25}
H_{HL}\Big|_{m\alpha}=
\begin{pmatrix}
    0 & -\Lambda^*_{m\alpha}(k)\Delta \\
    \Lambda_{m\alpha}(-k)\Delta & 0
\end{pmatrix}.
\end{eqnarray}
The relevant subspace is $mp\uparrow,\, mh\downarrow \times \alpha p\uparrow,\,
\alpha h\downarrow$. We consider triplet pairing with net $S_z=0$ and adopt the
convention that $\uparrow$ corresponds to the particle sector while $\downarrow$
corresponds to the hole sector. Combining Eqs.~\ref{S23}, \ref{S24}, and
\ref{S25}, we obtain the wave-function component in BM band $m$:
\begin{equation}
\psi_m(E)
=
\sum_{\alpha}
\begin{pmatrix}
0 &
\dfrac{\Lambda_{m\alpha}^*(k)\Delta}{\xi_m(k)}
\\[4pt]
\dfrac{\Lambda_{m\alpha}(-k)\Delta}{\xi_m(-k)} &
0
\end{pmatrix}
\psi_\alpha(E).
\label{S26}
\end{equation}
Here $\psi_m$ is the component of the wavefunction in the $H$-subspace
$mp\uparrow, mh\downarrow$, and similarly $\psi_\alpha$ is the component in the
$L$-subspace $\alpha p\uparrow, \alpha  h\downarrow$.

We next derive the flat-band components $\psi_\alpha(E)$. In the low-energy
sector, the Hamiltonian can be reduced to the $2\times2$ problem involving the
$H_{12}$ and $H_{21}$ couplings. In particular,
\begin{equation}
    H_{\alpha \beta}(\mathbf{k})=\delta\xi_{\alpha \beta}(\mathbf{k})+
    \begin{pmatrix}
    \bar{\xi}_{\alpha \beta}(\mathbf{k})  & \Lambda_{\alpha \beta}(\mathbf{k}) \Delta \\
    \mathrm{h.c.} & -\bar{\xi}_{\alpha \beta}(-\mathbf{k})
    \end{pmatrix},
    \label{1}
\end{equation}
where $\delta\xi_{\alpha \beta}(\mathbf{k})=[\xi_\alpha(\mathbf{k})-\xi_\beta(-\mathbf{k})]/2$,
$\bar{\xi}_{\alpha \beta}(\mathbf{k})=[\xi_\alpha(\mathbf{k})+\xi_\beta(-\mathbf{k})]/2$. The corresponding
eigenvalues are
$E_{\alpha,\pm}(\mathbf{k})=\delta\xi_{\alpha\beta}(\mathbf{k})\pm
\epsilon_\alpha(k)$ with $\epsilon_\alpha(k)=\sqrt{\bar{\xi}_{\alpha\beta}(\mathbf{k})^2+\big|\Lambda_{\alpha\beta}(\mathbf{k})\Delta\big|^2}$. The subindex $\alpha \beta$ indicates the $\alpha p \uparrow, \beta h \downarrow$ basis. The corresponding eigenvectors are
\begin{equation}
f_{\alpha \beta,+}(k)=
\begin{pmatrix}
u_{\alpha,k}\\
e^{-i\theta_k}v_{\alpha,k}
\end{pmatrix},
\qquad
   f_{\alpha \beta,-}(k)=
\begin{pmatrix}
-\,e^{i\theta_k}v_{\alpha,k}\\
u_{\alpha,k}
\end{pmatrix}, \label{S28}
\end{equation}
with, $\exp^{i\theta_k}=  {\Lambda_{\alpha \beta}(k) \Delta}/{|\Lambda_{\alpha \beta}(k) \Delta|}$, real $u_{\alpha,k},v_{\alpha,k}\ge0$ and $u_{\alpha,k}^2+v_{\alpha,k}^2=1$: 
\begin{equation}
\begin{aligned}
u_{\alpha,k}^2
&=
\frac{1}{2}
\left(
1+\frac{\bar{\xi}_{\alpha\beta}(k)}
{\epsilon_\alpha(k)}
\right), \, 
v_{\alpha,k}^2
&=
\frac{1}{2}
\left(
1-\frac{\bar{\xi}_{\alpha\beta}(k)}
{\epsilon_\alpha(k)}
\right),
\end{aligned}
\label{4}
\end{equation}
 and $u_{\alpha,k}v_{\alpha,k}
=
 {
\left|\Lambda_{\alpha\beta}(k)\Delta\right|
}/{
2\epsilon_\alpha(k)
}.$ Note that $f$ in Eq.~\ref{S28} is written in the $\alpha\beta$ subspace. We now
connect it back to the full flat-band basis
$1p\uparrow, 1h\downarrow,2p \uparrow, 2h\downarrow$, which is the basis used
in Eq.~\ref{S26}. For example, for $E_{1,+}$ we read off the corresponding
wavefunctions in the full flat-band basis:
\begin{eqnarray}
   && \psi_L (E_{1,+})=(   u_{1,k},0, 0, v_{1,k}  ),\quad \psi_L (E_{1,-})=(   -v_{1,k},0, 0, u_{1,k}  ) \nonumber \\
  &&  \psi_L (E_{2,+})=(   0, v_{2,k} , u_{2,k}, 0   ), \quad \psi_L (E_{2,-})=(   0, u_{2,k} , -v_{2,k}, 0   ).  \nonumber \\ \label{S30}
\end{eqnarray}  
Therefore, combining Eqs.~\ref{S26} and \ref{S30}, we obtain the full BdG
wavefunction $\psi$ in the BM-band basis for the active BdG eigenstates.

\subsection{Geometric Contribution to the Current Matrix Elements}
We are now ready to derive an explicit expression for the current matrix elements
in the active BdG bands:
\begin{align}
&\langle \varphi_i^p(k)|
J(\mathbf{k})
|\varphi_j^p(k)\rangle
={} 
\sum_{m\alpha}
[\varphi_i^p(k)]_\alpha^*
J_{\alpha m}
[\varphi_j^p(k)]_m
\nonumber\\
&+
\sum_{m\alpha}
[\varphi_i^p(k)]_m^*
J_{m\alpha}
[\varphi_j^p(k)]_\alpha
 +
[\alpha\alpha,mm] .
\end{align}
Here $\alpha$ is the flat-band index while $m$ is the remote-band index. The
terms in the bracket denote the flat--flat and remote--remote contributions,
which we do not consider here. The flat--flat contribution is suppressed by the
flat-band width, which is small in the narrow-band limit. The remote--remote
contribution is suppressed by $1/M^2$, where $M$ is the energy separation
between the flat and remote bands. (If the flat-band width is not small, the
flat--flat term is no longer negligible.)

From Eq.~\ref{S26}, we obtain
\begin{align}
J^p_{ij}
={}&
\Delta\,
\langle \varphi_i^p|
J\xi^{-1}\Lambda^*
|\psi_j^h\rangle_F
 +
\Delta\,
\langle \varphi_i^h|
\Lambda^T\xi^{-1}J
|\psi_j^p\rangle_F ,
\label{eq:Jp}
\end{align}
Here $J  \xi^{-1}\Lambda^*$ is a $2\times 2$ matrix in the flat-band
subspace $F$, and the subscript indicates that the expectation value is taken
within this flat-band subspace. The superscripts $p/h$ denote the particle/hole
components of the BdG wavefunctions. 

For the active bands, we label the energies by $E_{\alpha,s}$ and identify
$i=(\alpha s)$ and $j=(\alpha's')$. In the flat-band BdG basis, we simply take
$|\varphi_i \rangle_L=\psi_L(E_{\alpha,s})$ from Eq.~\ref{S30}. Using the general
relation for interband current matrix elements,
$J_{mn}=i(\xi_m-\xi_n)A_{mn}$, we arrive at 
\begin{align}
J^p_{ij}
={}&
-i\Delta
\Big[
(A\Lambda^*)_{\alpha\beta'}\,
\psi^*_{\alpha p}(E_{\alpha,s})\,
\psi_{\beta' h}(E_{\alpha',s'})
\nonumber\\
&\qquad
-(\Lambda^T A)_{\beta\alpha'}\,
\psi^*_{\beta h}(E_{\alpha,s})\,
\psi_{\alpha' p}(E_{\alpha',s'})
\Big].
\label{S33}
\end{align}
Here we use the wavefunctions in Eq.~\ref{S30}, and the subscripts specify the
corresponding basis components in $L$. Equation~\ref{S33} shows explicitly how
the Berry connection $A$ enters the current matrix elements, providing a
microscopic mechanism for the geometric structure of the active BdG bands.

The corresponding hole-sector current operator is
\begin{eqnarray}
   \tilde{J}^h_{ji}(-k)= \Delta  \langle \varphi_j^h | (J^T \xi^{-1}\Lambda)(-k)  | \psi_i^p\rangle + \nonumber \\  \Delta  \langle \varphi_j^p | (\Lambda^\dagger \xi^{-1} J^T    )(-k)  | \psi_i^h\rangle.
\end{eqnarray} 
 With $J^p$ and
$\tilde{J}^h$ in hand, one can obtain the active--active contribution.
Importantly, for all allowed combinations of $i,j$, the contribution is
geometric.

Equation~\ref{S33} makes this result explicit: the active-band current vertex is controlled by the product
of the pairing form factor and the Berry connection, rather than by the small bare velocity of the
flat bands alone.

\subsection{Geometric BFS}
A special case of the active--active channel is the intraband contribution,
$i=j$. The superconductor becomes gapless when
\begin{eqnarray}
   E_{1,+}(k)=-E_{2,-}(-k)=0 .
\end{eqnarray}
Here we label the flat bands such that $\xi_1(k) \leq \xi_2(k)$. The gapless
bands contribute to the superfluid stiffness through
\begin{eqnarray}
[\text{Gapless}]=-4\sum_{i,k} \frac{dn_F(E)}{dE}\Bigg|_{E=E_i(k)}   J^p_{ii}(k) \, \tilde{J}^h_{ii}(-k) 
\end{eqnarray}
Considering the case $i=1+$ and applying Eq.~\ref{S33} to the intraband matrix
element, we find
\begin{eqnarray}
J^p_{ii}(k) =-i\frac{|\Lambda_{12}(k) \Delta|}{2\epsilon_1(k)} \sum_{m \in \text{remote}}   \Big[ A_{1m} \Lambda^*_{m2} \Delta- \Lambda_{m2} A_{m1} \Delta  \Big].
\end{eqnarray}
For the hole-sector current operator, we similarly obtain
\begin{eqnarray}
   \tilde{J}^h_{ii}(-k)=i \frac{|\Lambda_{12}(k) \Delta|}{2\epsilon_1(k)} \sum_{m \in \text{remote}}   \Big[ A_{m2}  \Lambda_{m1} \Delta  -   \Lambda^*_{m1} A_{2m}\Delta     \Big]   \nonumber
\end{eqnarray} 
The calculation for the other gapless band is analogous. In general, this
leads to 
\begin{align}
D_s^{\mathrm{BFS}}
={}&
4\sum_{\alpha s,{\bf k}}
\left.
\frac{d n_F(E)}{dE}
\right|_{E=E_{\alpha,s}(k)}
\left(
\frac{
\Lambda_{\alpha\beta}(k)\Delta^2
}{
2\epsilon_\alpha(k)
}
\right)^2
 \\
&\times
\Gamma_{\alpha\beta}(\mathbf{k})
\Gamma_{\beta\alpha}(-\mathbf{k}). \nonumber
\end{align}
 Thus, the Bogoliubov Fermi surface is not merely a source of zero-energy density of states: its
gapless quasiparticles carry a quantum-geometric current vertex and therefore produce a finite
paramagnetic reduction of the superfluid stiffness.

\section{Active-Inactive Channel}
The static current-current correlation function is given by:
\begin{eqnarray}
Q_0(q=0)= -4\sum_{k} \sum_{ij}\frac{n_F(E_i)-n_F(E_j)}{E_i-E_j} \nonumber
  \\
  \times \langle \varphi^p_i(k) | J(\mathbf{k}) |\varphi^p_j(k) \rangle \, \langle \varphi^h_j(k) |  J^T(-\mathbf{k})|\varphi^h_i(k) \rangle.
\end{eqnarray}
We consider the contribution where one band belongs to the active band (labeled $\alpha$) and the other belongs to an inactive band (labeled $m$). This includes both the ($i=\alpha, j=m$) and ($i=m, j=\alpha$) terms, which contribute equally by symmetry.

For the remote bands, the typical interband energy splitting (on the order of $W \approx 50$ meV) is much larger than the superconducting gap, except near a few band-crossing points. Consequently, pairing between different remote bands is energetically unfavorable. We therefore adopt an intraband-pairing approximation for the BdG wavefunctions in the inactive sector. The resulting intraband Hamiltonian for a remote band $m$ reads:
\begin{eqnarray}
h_m(k)=    \begin{pmatrix}
        \xi_m(k) & \Lambda^*_{mm}\Delta \\
        h.c. &  -\xi_m(-k)
    \end{pmatrix}.
\end{eqnarray}
This Hamiltonian possesses a standard BCS structure. Its eigenvalues are $E_{m,\pm}(k)=[\xi_m(k)-\xi_m(-k)]/{2}\ \pm \epsilon_m$ with
\begin{equation}
 \epsilon_m=\sqrt{\bar\xi_m(k)^2+|\Lambda_{mm}(k)\Delta|^2} ,
\end{equation}
and $\bar\xi_m(k)=[\xi_m(k)+\xi_m(-k)]/{2}$. The corresponding eigenvectors are given by
\begin{equation}
f_{m,+}(k)=
\begin{pmatrix}
u_{m,k}\\
e^{-i\theta_k}v_{m,k}
\end{pmatrix},
\, 
f_{m,-}(k)=
\begin{pmatrix}
-\,e^{i\theta_k}v_{m,k}\\
u_{m,k}
\end{pmatrix},\,  
\end{equation}
where  $\exp{i\theta_k}=  {\Lambda^*_{mm}(k) \Delta}/{ |\Lambda_{mm}(k) \Delta|}$, the coherence factors are real, $u_{m,k},v_{m,k}\ge0$, satisfying $u_{m,k}^2+v_{m,k}^2=1$:
\begin{equation}
u_{m,k}^2=\frac12\left(1+\frac{\bar\xi_m(k)}{\epsilon_m(k)}\right),
\qquad
v_{m,k}^2=\frac12\left(1-\frac{\bar\xi_m(k)}{\epsilon_m(k)}\right)
\end{equation}
and 
$u_{m,k}v_{m,k}={|\Lambda_{mm}(k) \Delta|}/{2\epsilon_m(k)}$.

We now return to the product of current operators in the response function. Considering the specific case where $i=\alpha$ (flat) and $j=m$ (remote), the involved wavefunction overlaps are:
$
s u_{\alpha,k} v_{\alpha,k}\times  s' u_{m,k} v_{m,k}.
$
Here we have absorbed phases into field operators of each BM band wavefunction. Substituting these into the expression for the stiffness, and multiplying by a factor of 2 to account for the symmetric $i \leftrightarrow j$ contribution, the total active-remote contribution reads:
\begin{eqnarray}
 D_s^{\text{Remote}} =-8 \sum_{k,mss'\alpha}  ss'\frac{n_F(E_{\alpha,s})-n_F(E_{m,s'})}{E_{\alpha,s}-E_{m,s'}} \nonumber \\
  \times u_{\alpha,k} v_{\alpha,k} u_{m,k} v_{m,k}  J_{\alpha m} J_{\beta m}(-k).
\end{eqnarray}
Using the approximation $J_{\alpha m} = i (\xi_\alpha -\xi_m) A_{\alpha m} \approx -i \xi_m A_{\alpha m}$, and retaining the leading contribution from terms where $s=-s'$, the remote part is given by 
 \begin{eqnarray}
 D_s^{\text{Remote}} \simeq  2\sum_{k,ms\alpha}  s [n_F(E_{m,-s})-n_F(E_{\alpha,s})] \nonumber 
  \\ 
  \times \frac{ |\Lambda_{mm}(k)\Lambda_{\alpha \beta}(k) \Delta^2|}{\epsilon_\alpha}\times   A_{\alpha m} A_{\beta m}(-k).
  \label{eq:Ds_remote_final}
\end{eqnarray}

To elucidate the geometric structure of this result, we introduce a Berry connection in Nambu space,
\begin{eqnarray}
a_{\alpha m}(k)= i\langle U_\alpha|\nabla|U_m\rangle
= A_{\alpha m}(k)+A_{\beta m}(-k).
\end{eqnarray}
Here, $U_\alpha=[u_\alpha(k),u^*_\beta(-k)]$ is an eigenvector in the active sector, while $U_m=[u_m(k),u_m^*(-k)]$ is the BdG wavefunction for the $m$-th remote band within the intraband-pairing approximation. With this definition, we find the identity:
\begin{eqnarray}
   2A_{\alpha m}(k)A_{\beta m}(-k)
= a_{\alpha m}^2(k)-A_{\alpha m}^2(k)-A_{\beta m}^2(-k).
\end{eqnarray}
Consequently, $D^{\text{Remote}}_s$ can be expressed as a functional of the Cooper-pair wavefunction and the Berry connections defined in both the normal (BM) band basis and Nambu space. In contrast to the Kekul\'e case, for time-reversal-symmetric pairing processes, the Nambu-space connection $a_{\alpha m}$ vanishes.

If we further assume that the pairing potential $\Lambda_{mm}$ is approximately uniform across the remote bands, then Eq.~(\ref{eq:Ds_remote_final}) at zero temperature is proportional to a combination of metric tensors. Recognizing that the connections $A$ and $a$ are purely imaginary, their squares sum to the negative of their respective metrics ($A^2 \to -g$). Thus, the identity yields:
\begin{eqnarray}
 \sum_m   2A_{\alpha m}(k)A_{\beta m}(-k)=g_\alpha(k)+g_\beta(-k)-G_\alpha(k),
\end{eqnarray}
where $g_\alpha= \langle \nabla u_\alpha |\mathcal{P}| \nabla u_\alpha \rangle$ and $G_\alpha= \langle \nabla U_\alpha |\mathcal{P}'|  \nabla U_\alpha \rangle$. Here, $\mathcal{P}= \sum_{m \in \text{remote}}|  u_m \rangle \langle u_m | $ and $\mathcal{P}'=\sum_{m \in \text{remote}}|  U_m \rangle \langle U_m| $ are the projection operators onto the remote bands in the normal and Nambu spaces, respectively.

\section{Acknowledgement}
We thank Miuko Tanaka and Joel Wang for sharing their data with us. We also thank Enrico Rossi, Eduardo Fradkin and Kevin Nuckolls for helpful discussions.   

\section{Funding statement}
Q. C. is supported by the Innovation Program for Quantum Science and Technology (Grant No. 2021ZD0301904).
R.~B. is supported by the Department of Physics and Astronomy, Dartmouth College.
We also acknowledge the University of Chicago's Research Computing Center for their support of this work.

\section{Author Contributions}
K.L. conceived and supervised the project. K.W. performed the computations. K.W., Q.C. and R. B. contributed to the acquisition of the data and preparation of figures. All authors have contributed to the interpretation of the data and the drafting as well as the revision of the manuscript.

\section{Competing Interests}
The authors declare no competing interests.

\bibliography{ref}

\end{document}